\documentclass{emulateapj}
\usepackage{amsmath}
\usepackage{epsfig}
\usepackage{amssymb}
\usepackage{graphicx}
\usepackage{color}
\usepackage{natbib}
\usepackage{soul}

\usepackage{apjfonts} 
\usepackage{amsmath,amstext}
\usepackage[breaklinks,colorlinks,citecolor=blue,linkcolor=magenta]{hyperref} 
\usepackage[all]{hypcap} 
\usepackage{verbatim}

\usepackage{xcolor}

\begin{document}

\title{The lifetimes of high-redshift quasars suggest magnetic disk support}
\author{Jarrett Johnson\altaffilmark{1}, Phoebe Upton Sanderbeck\altaffilmark{1}, Nicole Lloyd-Ronning\altaffilmark{1}, Madeline Marshall\altaffilmark{1,2} and Kelcey Davis\altaffilmark{1,3,4}}

\altaffiltext{1}{Los Alamos National Laboratory, Los Alamos, NM 87545, USA}
\altaffiltext{2}{Oppenheimer Postdoctoral Fellow}
\altaffiltext{3}{Department of Physics, 196A Auditorium Road, Unit 3046, University of Connecticut, Storrs, CT 06269, USA}
\altaffiltext{4}{National Science Foundation (NSF) Graduate Research Fellow}

\begin{abstract}
It has recently been suggested that a variety of data on active galactic nuclei (AGN) can be explained if AGN disks are supported against gravitational fragmentation by magnetic fields that are advected into the disk from the surrounding galaxy.  Here we derive the maximum timescales over which accretion onto a black hole (BH) powering an AGN can be maintained at a given rate, both with and without magnetic disk support.  We then compare these timescales to the lifetimes of episodes of sustained luminous accretion that are inferred from measurements of the photoionized proximity zones around high-redshift quasars.  While some of the shortest inferred quasar lifetimes are consistent with pure gas pressure support, we find that some additional magnetic support is likely required to explain the longest inferred quasar lifetimes of > 10$^4$ yr. {\color{black} For these longest-lived AGN, we find that magnetic pressure in their disks can be up to a hundred times higher than the gas pressure}. In addition, the lack of inferred quasar lifetimes that are definitively > 10$^6$ yr is consistent with gas pressure and advected magnetic fields being the principal sources of disk support. This adds to the body of evidence that magnetic fields play an important role in sustaining the rapid growth of supermassive BHs in the early universe. 
\end{abstract}


\maketitle

\section{Introduction} 

It remains one of the foremost challenges in astrophysics and cosmology to explain how supermassive black holes (BHs) grew within the first billion years of cosmic history at $z$ $\ge$ 6 \citep{volonteri12, Johnson2016, inayoshi20}. Some new data from the JWST suggest small quasar duty cycles \citep{Pizzati2024, Huang2026, MengZhang2026}, which implies that BH growth in high-$z$ active galactic nuclei (AGN) is either largely obscured \citep{Gilli2022,JUS2022,Satya2023b} or that it occurs episodically, via a series of periods of sustained high rates of accretion.  The lengths of these episodes of sustained accretion provide critical constraints on BH growth models.  

Measurements of the sizes of so-called proximity zones around quasars during the epoch of reionization provide evidence for a large range in the durations of episodes of unobscured accretion for the BHs in AGN, from of order 10$^2$ to of order 10$^5$ years \citep{Eilers2021, Satyavolu2023,Duro2025}.  While these timescales are  consistent with episodic luminous accretion driving the growth of supermassive BHs during reionization, it remains an open question what determines these lifetimes and, in particular, whether there is an upper limit to the duration of an episode of luminous accretion.

It has been suspected for some time that magnetic fields produced by the magnetorotational instability (MRI) can support AGN accretion disks against gravitational fragmentation, thereby allowing for higher sustained accretion rates than would be possible in the absence of magnetic fields \citep{PP2005, BegPring2007, BA2023,Tsung2025,GD2025}.  Recently, \citet{Hopkins2024analytic, Hopkins2024fluxfreeze} have shown that the magnetic fields from the general interstellar medium (ISM) that are advected with the accretion flow can support sustained accretion even at highly super-Eddington rates.  In line with this, \citet{Hopkins2026maser} argue that observations of masers and AGN broad-line regions suggest that AGN disks are likely supported by magnetic pressure at their outskirts.  With the additional inference of high magnetic field strengths of $>$ 10$^4$ G near the inner edges of many AGN disks \citep{Daly2019}, it appears that magnetic support may play a role in supporting AGN disks at all scales.

{\color{black}
Here we estimate the timescales over which sustained accretion can be maintained in AGN disks, both with and without magnetic pressure support.  We build on the previous theoretical estimate of the duration of luminous AGN accretion presented by \citet{KN2015}.  In response to evidence for AGN flickering on timescales \citep{Schawinski2015}, these authors estimated a timescale of $\sim$ 10$^5$ yr for the duration of luminous accretion episodes.  While this study assumed a very thin accretion disk, it did not consider the role the magnetic fields may play in supporting AGN disks against fragmentation. Here we relax the assumption of a very thin accretion disk and explore the role of magnetic support of the accretion flow, which has been shown to result in thicker, magnetically elevated disks \citep{Sad2016, BA2023}. In turn, we find that the inclusion of magnetic field support in our estimate of the maximum duration of accretion episodes is key to explaining recent data on the lifetimes of high-z quasars.}

In the following section we derive analytic formulae for these timescales and in Section 3 we compare them to the inferred lifetimes of quasars.  In Section 4 we close with our conclusions and a brief discussion.

\section{Derivations}\label{sec:derivation}
We present two estimates of the maximum time for which accretion onto a BH of mass $M$ can be maintained at a given rate $\dot{M}$.  The first is derived under the assumption that gas pressure supports the disk against gravitational fragmentation, and the second relaxes this assumption to allow for the possibility that magnetic fields in the disk also contribute to its stability.

\subsection{Pure gas pressure support}
We begin by defining the timescale $t_{\rm acc}$ on which gas in the disk that starts at a distance $r$ from the BH is accreted onto the BH: 

\begin{equation}
\label{A}
t_{\rm acc} \simeq \frac{r}{v_{\rm r}} \mbox{ .}
\end{equation}
In the standard $\alpha$ disk formalism \citep{SS1973} the radial infall velocity is defined as

\begin{equation}
\label{B}
v_{\rm r} = \frac{3}{2}\frac{\alpha H^2 \Omega}{r} \mbox{ ,}
\end{equation}
where $\alpha$ is the dimensionless viscosity parameter, $\Omega = (GM/r^3)^{\frac{1}{2}}$ is the Keplerian angular velocity in the disk, and the disk scale height is defined in terms of the sound speed $c_{\rm s}$ of the gas as $H$ = $c_{\rm s}$/$\Omega$.  Here $G$ is Newton's constant.

The sound speed can be expressed in terms of the temperature $T$ of the gas in the disk as $c_{\rm s}^2$ = 3$k_{\rm B}$$T$/$\mu$$m_{\rm H}$, where $\mu$ is the mean molecular weight, $m_{\rm H}$ is the mass of the hydrogen atom, and $k_{\rm B}$ is Boltzmann's constant.  In turn, the temperature of the gas can be estimated under the standard assumption that it radiates energy away at the same rate that it gains energy as it falls toward the BH:

\begin{equation}
\label{C}
T = \left(\frac{3}{8 \pi} \frac{G M \dot{M}}{\sigma_{\rm SB} r^3}     \right)^{\frac{1}{4}} \mbox{ ,}
\end{equation}
where $\sigma_{\rm SB}$ is the Stefan-Boltzmann constant. Noting that Equation~\ref{C} implies that the sound speed in the disk drops with distance $r$ from the BH, we adopt the following expression for the maximum rate at which gas can accrete within $r$ without the disk fragmenting under its own gravity \citep{BegPring2007, Zhu2012, KN2015}:

\begin{equation}
\label{D}
\dot{M} = \frac{3 \alpha c_{\rm s}^3}{G} \mbox{ .}
\end{equation}
{\color{black} This expression follows from the relation $\dot{M}$=2$\pi$$r$$v_{\rm r}$$H$$\Sigma$ and from the condition of marginal gravitational stability given by $Q$ = 1, where $Q$ = $c_{\rm s}$$\Omega$/$G$$\pi$$\Sigma$ is the Toomre parameter.}

Using Equations~\ref{B}, \ref{C}, and \ref{D} in Equation~\ref{A}, we arrive at the following expression for the maximum time over which accretion onto the BH at a given rate can be maintained:

\begin{equation}
\label{E}
t_{\rm acc} = 2 \left( 3 \alpha \right)^{\frac{1}{9}} \left( \frac{3 k_{\rm B}}{\mu m_{\rm H}} \right)^{\frac{2}{3}} \left( \frac{3}{8 \pi \sigma_{\rm SB}}  \right)^{\frac{1}{6}} \frac{M^{\frac{2}{3}}}{G^{\frac{4}{9}} \dot{M}^{\frac{17}{18}}}   \mbox{ .}
\end{equation}
Assuming $\mu$ = 0.6, as appropriate for an ionized gas, this is 

\begin{equation}
\label{F}
t_{\rm acc} \simeq 10^4 \, {\rm yr} \, \left(\frac{\alpha}{0.1}  \right)^{\frac{1}{9}} \left(\frac{M}{10^9 \, {\rm M}_{\odot}  }  \right)^{\frac{2}{3}}  \left(\frac{\dot{M}}{100 \, {\rm M}_{\odot} {\rm yr}^{-1} }  \right)^{-\frac{17}{18}}  \mbox{ .}
\end{equation}
As we will show in Section~\ref{sec:data}, this timescale is in line with the lifetimes inferred for many high-redshift quasars, but it does not appear to be consistent with the longest such lifetimes.

\subsection{Magnetic and gas pressure support}
To account for the additional support of the disk by magnetic fields, we follow previous work \citep{KimOst2001, GD2025, Tsung2025} in adopting the following modified form of the scale height:

\begin{equation}
\label{G}
H = \frac{\sqrt{c_{\rm s}^2 + v_{\rm A}^2}}{\Omega} \mbox{ ,}
\end{equation}
where $v_{\rm A}$ is the Alfv{\'e}n speed. Accordingly, we also adopt the following modified form for the maximum accretion rate:

\begin{equation}
\label{H}
\dot{M} = \frac{3 \alpha}{G} \left(c_{\rm s}^2 + v_{\rm A}^2  \right)^{\frac{3}{2}} \mbox{ .}
\end{equation}
{\color{black}Like Equation~\ref{D}, this follows from the condition of marginal gravitational stability given by $Q$ = 1, but with $H$ as defined in Equation~\ref{G} and with $Q$ defined as $Q$ = ($c_{\rm s}^2$+$v_{\rm A}^2$)$^{\frac{1}{2}}$$\Omega$/$G$$\pi$$\Sigma$, following \citet{KimOst2001}.} With these two modified equations, the maximum time for which accretion at a rate $\dot{M}$ can be maintained is

\begin{equation}
\label{I}
t_{\rm acc,mag}  =  \frac{2}{3^{\frac{1}{3}}} \left( \frac{3 k_{\rm B}}{\mu m_{\rm H}} \right)^{\frac{2}{3}} \left( \frac{3}{8 \pi \sigma_{\rm SB}}  \right)^{\frac{1}{6}} \frac{M^{\frac{2}{3}}}{\alpha^{\frac{1}{3}} c_{\rm s}^{\frac{4}{3}} \dot{M}^{\frac{1}{2}}} \mbox{ .}
\end{equation}
Again assuming $\mu$ = 0.6 and expressing the sound speed $c_{\rm s}$ in terms of gas temperature $T$, this becomes 

\begin{equation}
\label{J}
\begin{split}
t_{\rm acc,mag}  \simeq 10^5 \, {\rm yr} \, \left(\frac{\alpha}{0.1}  \right)^{-\frac{1}{3}} \left(\frac{M}{10^9 \, {\rm M}_{\odot}  }  \right)^{\frac{2}{3}}  \\
  \left(\frac{\dot{M}}{100 \, {\rm M}_{\odot} {\rm yr}^{-1} }  \right)^{-\frac{1}{2}} \left( \frac{T}{10^4 \, {\rm K}} \right)^{-\frac{2}{3}} 
\mbox{ .}
\end{split}
\end{equation}

\begin{deluxetable*}{lcccccccc}
\setlength{\tabcolsep}{9pt}
\tablecaption{Measured and calculated properties of high-redshift ($z$ > 5.8) quasars}
\tablehead{\colhead{Object} & $z$ & \colhead{$M$ [M$_{\odot}$]} & \colhead{$\dot{M}$ [M$_{\odot}$ yr$^{-1}$] } & \colhead{$\dot{M}/\dot{M_{\rm Edd}}$} & \colhead{$t_{\rm acc}$ [yr] } & \colhead{$t_{\rm acc,mag}$ [yr]}  & \colhead{$t_{\rm Q}$ [yr] } & \colhead{$\beta$}}
\startdata
         VDES J0330-4025  & 6.23 & 4.96 $\times$ 10$^9$    & 15.3 & 0.14    & 2.59 $\times$ 10$^5$ & 7.48 $\times$ 10$^5$ & 1.26 $\times$ 10$^4$ & 0.17 \\
         VDES J0323-4701  & 6.24 & 2.8 $\times$ 10$^8$     & 10.8 & 1.76    & 5.22 $\times$ 10$^4$ & 1.29 $\times$ 10$^5$ & 3.98 $\times$ 10$^5$ & 0.23 \\
         CFHQS J2100-1715 & 6.08 & 2.18 $\times$ 10$^9$    & 7.19 & 0.15    & 3.04 $\times$ 10$^5$ & 6.29 $\times$ 10$^5$ & 2.00 $\times$ 10$^2$ & 0.34  \\
         PSO J261+19      & 6.48 & 4.7 $\times$ 10$^8$     & 8.26 & 0.8    & 9.53 $\times$ 10$^4$ & 2.10 $\times$ 10$^5$ & 5.01 $\times$ 10$^5$  & 0.29 \\
         PSO J011+09      & 6.44 & 1.39 $\times$ $10^9$    & 22   & 0.72    & 7.82 $\times$ 10$^4$ & 2.66 $\times$ 10$^5$ & 1.00 $\times$ 10$^5$ & 0.13 \\
         SDSS J1143+3808  & 5.83 & 3.6 $\times$ 10$^{10}$  & 455  & 0.58    & 3.95 $\times$ 10$^4$ & 5.17 $\times$ 10$^5$ & 6.31 $\times$ 10$^4$ & 0.014 \\
         SDSS J0100+2802  & 6.32 & 1.04 $\times$ 10$^{10}$ & 222  & 0.97    & 3.40 $\times$ 10$^4$ & 3.23 $\times$ 10$^5$ & 1.26 $\times$ 10$^5$ & 0.024 \\
         PSO J158-14      & 6.07 & 1.58 $\times$ 10$^9$    & 29.5 & 0.85    & 6.47 $\times$ 10$^4$ & 2.51 $\times$ 10$^5$ & 6.31 $\times$ 10$^3$ & 0.1 \\
         SDSS J1335+3533  & 5.93 & 4.16 $\times$ 10$^9$    & 60.5 & 0.66    & 6.28 $\times$ 10$^4$ & 3.35 $\times$ 10$^5$ & 1.00 $\times$ 10$^3$ & 0.059\\
         CFHQS J2229+1457 & 6.15 & 1.44 $\times$ 10$^9$    & 2.2  & 0.07    & 6.99 $\times$ 10$^5$ & 8.58 $\times$ 10$^5$ & 7.94 $\times$ 10$^2$ & 1.9 \\
 \enddata
\tablecomments{Timescales $t_{\rm acc}$ and $t_{\rm acc,mag}$ are calculated assuming $\alpha$ = 0.1 and $T$ = 10$^4$ K, plasma $\beta$ parameters assuming $\alpha$ = 0.1, and Eddington fractions $\dot{M}$/$\dot{M_{\rm Edd}}$ assuming a radiative efficiency of $\epsilon$ = 0.1. We adopt the lifetimes $t_{\rm Q}$ presented in \citet{Eilers2021}, \citet{Duro2025}, and \citet{Duro2025b} for quasars for which are available proximity zone measurements \citep{Eilers2020}, as well as BH mass and accretion rate measurements \citep{Farina2022, Wu2022}.}
\label{table}
\end{deluxetable*}

We have retained the form of Equation~\ref{C} in deriving $t_{\rm acc,mag}$, implicitly assuming that the energy in the magnetic field is not drawn from the gravitational potential energy of the BH, but instead that the gravitational potential energy is predominantly converted into thermal energy.  This is in line with recent results suggesting that the magnetic fields in AGN disks may be largely sourced from the surrounding ISM and that the magnitude of the magnetic field is enhanced in accord with flux-freezing as the density of the gas increases toward the BH \citep{Hopkins2024analytic, Hopkins2024fluxfreeze}.
{\color{black} Consistent with this, \citet{Hopkins2024analytic} show that the alpha-disk temperature scaling we adopt matches the scaling produced by their analytical model for a disk with magnetic field supplied by advection from the ISM.}

{\color{black}  The degree of magnetic support in a disk can be characterized by the plamsa $\beta$ parameter, which is defined as the ratio of the gas pressure to the magnetic pressure.
Noting that the gas pressure is given by $P_{\rm gas}$ = $\rho$$k_{\rm B}$$T$/$\mu$$m_{\rm H}$ and that the magnetic pressure is given by $P_{\rm mag}$ = $B^2$/$8\pi$, where $B$ is the magnetic field strength, this implies that }  

{\color{black}
\begin{equation}
\label{JJ}
 \beta = \frac{2}{3} \frac{c_{\rm s}^2}{v_{\rm A}^2} \mbox{ .}
\end{equation}
}

{\color{black} Using Equation~\ref{H} to solve for $c_{\rm s}$ in terms of $\beta$, and then substituting into Equation~\ref{I}, we find the following alternative expression for the maximum duration of sustained accretion through a magnetized disk:}

{\color{black}
\begin{equation}
\label{LL}
t_{\rm acc,mag} = 2 \left( 3 \alpha \right)^{\frac{1}{9}} \left( \frac{3 k_{\rm B}}{\mu m_{\rm H}} \right)^{\frac{2}{3}} \left( \frac{3}{8 \pi \sigma_{\rm SB}}  \right)^{\frac{1}{6}} \frac{M^{\frac{2}{3}}}{G^{\frac{4}{9}} \dot{M}^{\frac{17}{18}}} \left( \frac{2}{3 \beta} + 1 \right)^{\frac{2}{3}}  \mbox{ .}
\end{equation}
}

{\color{black} Using Equation~\ref{E}, this implies that the fractional increase in the maximum duration of sustained accretion that is due to magnetic pressure support, relative to that due to pure gas pressure support, is }

\begin{equation}
\label{LL}
\frac{t_{\rm acc,mag}}{t_{\rm acc}} = \left( \frac{2}{3 \beta} + 1 \right)^{\frac{2}{3}}    \mbox{ .}
\end{equation}
This shows that the impact of magnetic pressure support on the maximum duration of an accretion episode is determined directly by the ratio of the gas pressure to the magnetic pressure. 

The reason that magnetic pressure enables longer accretion episodes is that it stabilizes the disk out to larger radii than gas pressure alone, allowing for a steady accretion flow of gas from farther out in the disk.  In turn, because material from farther out takes more time to reach the BH, as shown by Equation~\ref{A}, this translates into longer episodes of steady accretion. 
{\color{black}
 Using Equation~\ref{B} for $v_{\rm r}$ in Equation ~\ref{A}, and using the scalings with the disk radius $r$ of all the quantities therein that we have adopted thus far, yields 

 \begin{equation}
\label{MM}
t_{\rm acc,mag} \propto \frac{r^{5/4}}{\alpha \left(1+\beta^{-2}\right)} \mbox{ .}
\end{equation}
This shows that, for a BH of a given mass accreting at a given rate, while the accretion timescale decreases with $\alpha$ and with the strength of the magnetic field through $\beta$, it nonetheless increases with the radius $r$ out to which the accretion disk is gravitationally stable.  
}

\section{Comparison to data}
\label{sec:data}
Here we test our derived upper limits for the duration of BH accretion against the inferred durations of luminous accretion onto high-redshift BHs, the so-called quasar lifetime $t_{\rm Q}$. We adopt the lifetimes presented in \citet{Eilers2021}, \citet{Duro2025}, and \citet{Duro2025b} for quasars for which are available proximity zone measurements \citep{Eilers2020}, as well as BH mass and accretion rate measurements \citep{Farina2022, Wu2022}.  This data is presented in Table \ref{table}.  Also shown in Table \ref{table} are the values of $t_{\rm acc}$ and $t_{\rm acc, mag}$ for each object.  These timescales are calculated using the measured BH masses and accretion rates, following Equations~\ref{F} and ~\ref{J} respectively, assuming  characteristic values of $\alpha$ = 0.1 and $T$ = 10$^4$ K.

\begin{figure}
\centering
\resizebox{9.2cm}{!}{\includegraphics{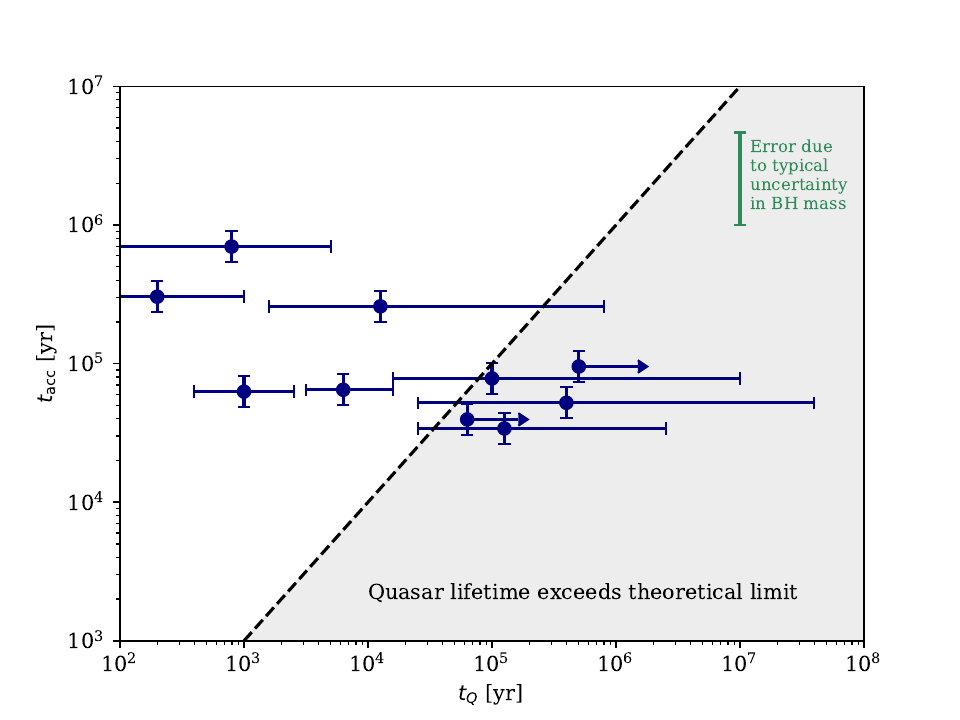}}\\
\caption{The theoretical maximum durations $t_{\rm acc}$ for which accretion can be maintained, for the case of pure gas pressure disk support, plotted against the observationally-inferred quasar lifetimes $t_{\rm Q}$, for the objects presented in Table \ref{table}. {\color{black} Data points lying to the right of the dashed line represent AGN disks for which $t_{\rm Q}$ > $t_{\rm acc}$, and which are thus unlikely to be supported by gas pressure alone}.  Note that the error bars on $t_{\rm acc}$ are small, due to the weak dependence of $t_{\rm acc}$ on $\alpha$ in Equation~\ref{F}.  The green error bar represents the additional uncertainty in $t_{\rm acc}$ due to the typical uncertainty in the BH mass $M$, following \citet{Satyavolu2023}.  }
\label{fig:no_mag}
\end{figure}

\begin{figure}
\centering
\resizebox{9.2cm}{!}{\includegraphics{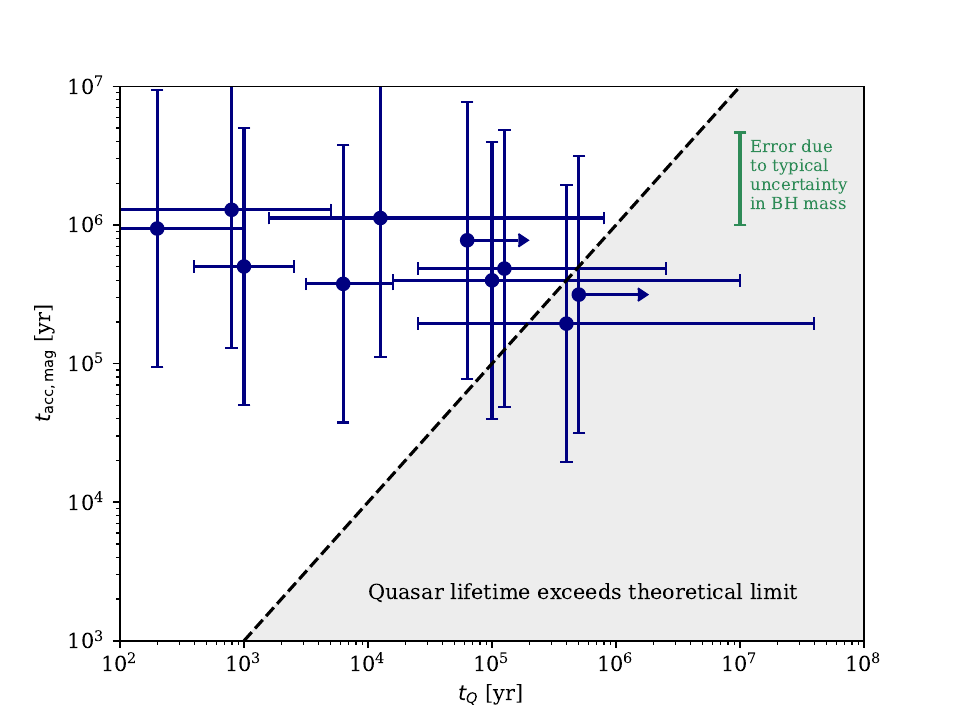}}\\
\caption{The theoretical maximum durations $t_{\rm acc,mag}$ for which accretion can be maintained, for the case of both gas and magnetic pressure disk support, plotted against the observationally-inferred quasar lifetimes $t_{\rm Q}$, for the objects presented in Table \ref{table}.  {\color{black} To within the measurement uncertainties, all of the data  lie to the left of the dashed line and are thus consistent with the AGN disks being supported by a combination of gas and magnetic pressure.} The green error bar represents the additional uncertainty in $t_{\rm acc, mag}$ due to the typical uncertainty in the BH mass $M$, following \citet{Satyavolu2023}. }
\label{fig:mag}
\end{figure}

For the uncertainties in these timescales that arise from the uncertainty in the disk temperatures in our sample of objects, we choose a range of 10$^3$ K $\le$ $T$ $\le$ 10$^5$ K, which brackets the temperatures that are typically found for AGN \citep{Zhang2024, Chen2026}.  It is also consistent with the gas temperatures in the AGN-candidate `Little Red Dots' (LRDs) that have been discovered using JWST \citep{Killi2024, Matthee2024, Naidu2025} and in AGN harboring some of the most massive BHs known at $z$ $\ga$ 8 \citep{Tripodi2024}. Indeed, this temperature range is also in line with measurements that have been carried out on galaxies up to $z$ $\simeq$ 11 \citep{Chakraborty2024, Alvarez2024}.  To evaluate the uncertainties in these timescales due to uncertainty in the values of $\alpha$ for the AGN in our sample, we adopt a range of 0.01 $\le$ $\alpha$ $\le$ 1. This brackets the range typically inferred for AGN disks \citep{King2007, Liu2008, Xue2025}, {\color{black} as well as roughly covering the range typically found in simulations of AGN disks \citep{salvesen2016, Jiang2019b, Mishra2020}}.  The vertical error bars on the data shown in Figures \ref{fig:no_mag} and \ref{fig:mag} reflect these ranges of possible values for $\alpha$ and $T$.  The horizontal error bars reflect the uncertainties in the values of the quasar lifetimes reported in the literature.

Figure \ref{fig:no_mag} presents a comparison between the quasar lifetime $t_{\rm Q}$ and the maximum accretion timescale $t_{\rm acc}$, derived assuming only gas pressure support in the disk, for the AGN listed in Table \ref{table}.  While the objects with the shortest inferred lifetimes $t_{\rm Q}$ $<$ 10$^4$ yr are consistent with our derived maximum accretion timescale, it is clear that there are objects with longer lifetimes approaching 10$^6$ yr that are not consistent with it.

Figure \ref{fig:mag} presents a comparison between the quasar lifetime $t_{\rm Q}$ and the maximum accretion timescale $t_{\rm acc, mag}$, derived assuming both gas and magnetic pressure support in the disk, for the same set of data.  In this case, none of the inferred quasar lifetimes exceed the theoretical maximum accretion timescale, to within the measurement uncertainties.  Based on the disagreement with the data presented in Figure \ref{fig:no_mag} and the agreement shown in Figure \ref{fig:mag}, we conclude that magnetic pressure support is likely responsible for the longest inferred lifetimes of high-redshift quasars. 

{\color{black} Given the timescales $t_{\rm acc}$ and $t_{\rm acc,mag}$ that we have calculated for our sample of quasars, we can use Equation~\ref{LL} to estimate the plasma $\beta$ parameters characterizing their accretion disks.  These estimated values for $\beta$, calculated assuming $\alpha$ = 0.1, are presented in Table 1. The values for $\beta$ span a range from roughly of order 10$^{-2}$ to 1, and are generally larger than the 10$^{-4}$ $<$ $\beta$ $<$ 10$^{-2}$ produced in the outskirts of the disk simulation presented in \citet{Hopkins2024fluxfreeze}. This may be due to the magnetic fields advected from the ISM being weaker in the host galaxies of the high-z quasars in our sample than assumed in this simulation.  Nonetheless, the fact that we find $\beta$ $<<$ 1 for most of the disks powering the high-z quasars in our sample is a clear indication that magnetic pressure likely plays a dominant role in supporting them against gravitational fragmentation.}

\section{Discussion and Conclusions}
\label{sec:conclusions}
We have derived the maximum timescales over which accretion onto a BH from a disk can be maintained at a given rate, accounting for both gas and magnetic pressure support of the disk against gravitational fragmentation.  Applying this to AGN, we have compared these calculated upper limits to the inferred lifetimes of luminous quasars during the epoch of reionization. Our results suggest that the quasars with the longest lifetimes are likely powered by accretion through AGN disks that are magnetically supported, {\color{black} with the magnetic pressure in some cases being roughly one hundred times higher than the gas pressure}. This is complementary to other recent work showing that magnetic support, particularly in the outer regions of AGN disks, may be required to explain observations of masers and  broad-line regions \citep{Hopkins2026maser}.  Our results also suggest that the shortest quasar lifetimes are consistent with predominantly gas pressure support, although they do not rule out some degree of magnetic support in these objects either.

 While we have estimated the maximum duration for which a given accretion rate can be maintained, we can also estimate the corresponding size of the AGN disk within which it can be maintained.  Equation~\ref{J} implies that gas and magnetic pressure support are sufficient to maintain steady accretion at a rate $\dot{M}$ for a time $t_{\rm acc,mag}$ through the region of the disk interior to that at which the gas temperature is $T$.  Equation~\ref{C} provides the following estimate for the location at which the gas temperature is $T$:

\begin{equation}
\label{K}
r  = \left(\frac{3 G M \dot{M}}{8 \pi \sigma_{\rm SB} T^4} \right)^{\frac{1}{3}}\mbox{ ,}
\end{equation}
which, again normalizing to typical values, is

\begin{equation}
\label{L}
r  \simeq 10^{17} \, {\rm cm} \left(\frac{M}{10^9 {\rm M}_{\odot}} \right)^{\frac{1}{3}} \left(\frac{\dot{M}}{100 \,  {\rm M}_{\odot} {\rm yr}^{-1}} \right)^{\frac{1}{3}} \left(\frac{T}{10^4 \, {\rm K}}\right)^{-\frac{4}{3}} \mbox{ .}
\end{equation}
This distance is consistent with the outer edges of dusty AGN disks that are inferred from reverberation mapping, particularly assuming that the gas temperatures are similar to the inferred dust temperatures of $\sim$ 10$^3$ K \citep{Landt2023}.

There is some additional data on the durations of luminous AGN accretion episodes to which we can compare our estimates. The AGN lifetime scaling with BH mass that \citet{Huse2022} infer from extended narrow-line region measurements, $t_{\rm Q}$ $\propto$ $M^{0.45}$,  is similar to the scaling we have derived,  $t_{\rm acc,mag}$ $\propto$ $M^{\frac{2}{3}}$. In addition, the timescale $t_{\rm acc,mag}$ $\simeq$ 10$^5$ yr that we have derived for the duration of magnetically-supported accretion, presented in Equation~\ref{J}, is also in line with the interpretation of AGN flickering on timescales of $\simeq$ 10$^5$ yr reported by \citet{Schawinski2015} and \citet{Comerford2017}, as well as with the range of 10$^4$--10$^6$ yr which is inferred from the Ly$\alpha$ forest observations of \citet{KirkTytler2008} and from the recent multi-wavelength observations presented by \citet{Dai2026}.  {\color{black} Finally, it is also consistent with the $\la$ 10$^6$ yr quasar lifetime limit recently inferred from relic ionized regions surrounding overmassive BHs at $z$ $\simeq$ 6.6 \citep{Meyer2026}.} The broad agreement with these independent measurements of the lifetimes of luminous AGN accretion episodes, in addition to that which we have found with high-redshift quasar lifetimes, suggests that magnetic AGN disk support may be relatively universal.

Other additional sources of AGN disk support that we have not included in our estimated disk lifetimes are MRI-generated magnetic fields, turbulence, and radiation pressure \citep{Jiang2019b, jiang2019}.  Any energy that goes into these would ultimately come from the gravitational potential energy of the BH, which would in turn mean less thermal energy available to support the disk than we have assumed in our estimates.  That said, unlike thermal energy, turbulent kinetic energy and magnetic energy generated via the MRI would also not be promptly radiated away, which means they could in principle provide additional support beyond what is provided by the thermal and advected magnetic energy that we have included in our estimates.  
However, \citet{Hopkins2024fluxfreeze} find that the magnetic fields advected from the ISM provide for higher Alfv{\'e}n speeds than are likely to be generated by the MRI, suggesting that magnetic fields amplified by the MRI are not likely to provide significant additional support of the gas against gravitational fragmentation.  Finally, we note that if there was another dominant form of disk support beyond what we have included in our estimates for the disk lifetime, we would expect there to be significantly longer disk lifetimes than the maximum values of order 10$^5$ yr that have so far been reported for reionization-era quasars.  If there are definitively longer disk lifetimes inferred from future observations of high-$z$ quasar proximity zones, then this would constitute evidence of additional sources of support beyond the gas pressure and advected magnetic fields that we have considered here.

In principle, an alternative explanation for the lifetimes of high-redshift quasars may be the duration over which gas is supplied to the AGN disk from the ISM of the host galaxy.  This would in turn demand an explanation as to why the gas supply from the ISM would be modulated on timescales that do not exceed the observed lifetimes of up to of order 10$^5$ yr.  Cosmological simulations have shown that stellar feedback in the ISM of AGN-hosting galaxies at $z$ $>$ 4 may limit the duration of episodes of BH accretion at the highest rates to $<$ 10$^6$ yr, but they also show that BH growth to the large masses of $>$ 10$^8$ M$_{\odot}$ characterizing our sample of SMBHs at $z$ $>$ 6 occurs during much longer periods over which gas is readily supplied to the central BH from the ISM \citep{Alcazar2017}.  This suggests that a steady supply of gas from the ISM of the host galaxy is required to grow the most massive BHs in the early universe and, in turn, that a limited gas supply is not likely to be the factor dictating the duration of episodes of their growth through luminous accretion.

Given the significantly longer timescales over which we have found that BH accretion can be sustained from magnetically supported disks, as compared to those from disks that are solely gas pressure supported, it may be the case that most supermassive BHs in the early universe are grown at the centers of magnetized disks.  In turn, if it is the case that the main source of the disk magnetism is advection from the general ISM in the AGN host galaxy, as suggested by \citet{Hopkins2024analytic, Hopkins2024fluxfreeze}, then the processes which generate the first magnetic fields \citep{McD2026} in early galaxies may also be prerequisites for the sustained rapid growth of their central BHs.

\section*{Acknowledgments}
Work at LANL was done under the auspices of the NNSA of the US Department of Energy. M.~A.~M. is supported  by LDRD Oppenheimer Postdoctoral Fellowship 20240752PRD1.  K.~D. acknowledges support from a NSF Graduate Research Fellowship award number 2040433, NSF CAREER-1945546, NASA JWST-GO-06368, and JWST-GO-05718. The authors are grateful to Hui Li, Tim Waters, Greg Salvesen, and Anna-Christina Eilers for helpful feedback and discussions, as well as to an anonymous reviewer for very constructive suggestions.


\bibliography{References}



\end{document}